\documentclass[twocolumn,showpacs,preprintnumbers,amsmath,amssymb]{revtex4-2} 
\usepackage{graphicx}
\usepackage{dcolumn}
\usepackage{bm}
\usepackage{tabularx}
\usepackage{ulem} 
\newcolumntype{M}{>{\centering\arraybackslash}m{1.85cm}}
\usepackage[export]{adjustbox}
\usepackage{float}
\usepackage{braket}
\usepackage{lipsum}
\usepackage{amsmath}
\graphicspath{{figure/}}
\usepackage{xcolor}   

\usepackage{caption}
\usepackage{subcaption}

\usepackage{hyperref}
\hypersetup{
    colorlinks=true, 
    linktoc=all,     
    urlcolor  = cyan,
    citecolor = blue,
    linkcolor=blue,  
}

\makeatletter
\newcommand{\colorcaption}[2][]{%
  \begingroup%
  \renewcommand{\@caption@fignum@sep}{ (Color online). }%
  \caption[#1]{#2}%
  \endgroup%
}
\makeatother
\newcommand{\orcid}[1]{\href{https://orcid.org/#1}{\hskip2pt\includegraphics[width=9pt]{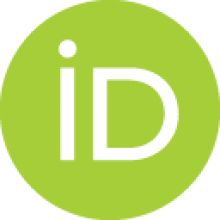}}}

\begin{document}

\title{  \textit{Ab initio} study of $\beta$-decay and pairing in $N=Z$ nuclei }

\author{Subhrajit Sahoo\orcid{0000-0001-8000-2150}}
\email{s$\_$sahoo@ph.iitr.ac.in}
\affiliation{Department of Physics, Indian Institute of Technology Roorkee, Roorkee 247667, India}

\author{Praveen C. Srivastava\orcid{0000-0001-8719-1548}}
\email{ praveen.srivastava@ph.iitr.ac.in}
\affiliation{Department of Physics, Indian Institute of Technology Roorkee, Roorkee 247667, India}

\date{\hfill \today}

\begin{abstract}

We investigate the $\beta$-decay properties of  $rp$-process waiting-point nuclei $^{72}$Kr, $^{68}$Se, and $^{64}$Ge from realistic nuclear forces based on chiral effective field theory. The \textit{ab initio} valence-space in-medium similarity renormalization group method is employed for this purpose to consistently derive Hamiltonians and Gamow-Teller operators from chiral two- and three-nucleon interactions. The calculated half-lives and branching ratios indicate that nearly the entire decay intensity is confined within 1 MeV of excitation energy in the daughter nuclei. We address the isoscalar and isovector pairing and their impact on ground state properties of these waiting-point nuclei, along with several other $N=Z$ systems in the $fp$-shell. Our results do not provide evidence for an isoscalar condensate or any dominant isovector pairing condensate-like phase in these $N=Z$ nuclei. We present the full $B(\mathrm{GT})$ strength distributions and discuss the influence of pairing correlations on them. The present work provides a microscopic picture of $\beta$-decay strengths and pairing in $N=Z$ nuclei far from the stability line.

\end{abstract}

\maketitle
\section{Introduction} \label{introduction}
Nuclear weak interaction processes are essential for understanding nucleosynthesis and the origin of elements in astrophysical environments \cite{LangankeReview, SuzukiReview, SchatzRep, Schatz2001}. For example, $\beta^+$-decay plays a key role in the rapid proton capture ($rp$) nucleosynthesis process, which is responsible for the synthesis of medium- to heavy-mass proton-rich nuclei ($A \sim 107$) \cite{Schatz2001, Wormer1994, Rubio2017}. In Type I x-ray bursts on accreting neutron stars, when $rp$-process advances towards the proton drip line, further proton captures become hindered, and the nucleosynthesis path is governed by relatively slow $\beta$-decays. This leads to the accumulation of several $N=Z$ nuclei, namely $^{64}$Ge, $^{68}$Se, $^{72}$Kr, $^{76}$Sr, etc., known as waiting-point (WP) nuclei. The $\beta^+$-decay and effective half-lives of these WP nuclei are crucial for determining and analyzing the reaction pathways and timescale, isotopic abundance patterns, as well as energy generation in x-ray bursts \cite{Parikh2013, Nacher2023}. In particular, $^{64}$Ge, the first waiting point encountered in the $rp$-process path, together with $^{68}$Se and $^{72}$Kr, play a decisive role in constraining and modeling x-ray bursts due to their comparatively long lifetimes \cite{Zhou2023, Chen2024}.

The $\beta$-decay or Gamow-Teller (GT) strength distributions of the parent nucleus to excited states of daughter nuclei not only determine half-lives but also encode essential information on the underlying nuclear structure.
The GT strength ($B(\mathrm{GT})$) distributions are sensitive to the ground state (g.s.) properties of the parent nucleus, such as deformation and spin, and have been used to probe shape evolution \cite{Hamamoto1995, Poirier2004, Nacher2004, Cerdan2009} and isospin symmetry breaking \cite{Hoff2020, Algora2025} along the $N \sim Z$ line--subjects that remain under active debate \cite{Waniganeththi2022, Wimmer2021, Lenzi2021, Sveiczer2022}. The $B(\mathrm{GT})$ distributions of WP nuclei are crucial inputs for astrophysical models used to compute stellar weak-interaction rates \cite{Chen2024, Sarriguren2009, Sarriguren2011, Jameel2012, Langanke2000, Langanke2001}. However, most WP nuclei lie close to the proton drip line, and their high-resolution $\beta$-decay measurements are particularly challenging. As a result, experimental information on GT strength distributions is typically constrained to low excitation energies and exists for a few WP nuclei. Measurements for $^{72}$Kr and $^{76}$Sr have been performed at CERN-ISOLDE \cite{Cerdan2009, Briz2015}, while studies of the decays of $^{64}$Ge and $^{68}$Se are presently underway \cite{Nacher2023}. In the absence of comprehensive experimental data, astrophysical calculations rely on inputs from theoretical models. This makes a microscopically grounded description of $B(\mathrm{GT})$ distributions essential to inform and constrain nucleosynthesis models and stellar decay rates, as well as to achieve reliable predictive power across the nuclear chart. These studies will be vital for interpreting current measurements and guiding future experiments at rare-isotope beam (RIB) facilities.

In this work, we present an \textit{ab initio} study of $\beta$-decay half-lives and $B(\mathrm{GT})$ distributions of the WP nuclei $^{64}$Ge, $^{68}$Se, and $^{72}$Kr in the $fp$-shell. These nuclei, with equal numbers of protons and neutrons, also provide ideal testing grounds for exploring neutron–proton ($np$) pairing correlations in addition to the conventional like-particle pairing. The near degeneracy of proton and neutron orbitals in the $N=Z$ nuclei allows the formation of both isovector ($J=0$, $T=1$) and isoscalar ($J=1$, $T=0$) $np$ pairs, together with the usual proton-proton and neutron-neutron pairs ($J=0$ and $T=1$). The identical-particle pairing influences the bulk properties of nuclei, such as binding energies, shell gaps, excitation energies, and angular momenta evolution \cite{PovesPLB1998, Brown2013, Dean2003, Tsunoda2020, Afanasjev2005}. They coexist on equal footing with isovector $np$ pairs in $N=Z$ nuclei, while the existence and role of isoscalar $np$ pairing continue to be subjects of intensive debate \cite{Frauendorf2014, Macchiavelli2000, Sagawa2013, Cederwall2011}. In addition to low-energy nuclear structure, the GT distributions are found to be sensitive to pairing correlations \cite{Sveiczer2022, Engel1997, Kaneko2018, Bai2013, Choudhary2025}. We therefore investigate how isovector and isoscalar pairing correlations manifest in the ground-state properties of the WP nuclei and study their impact on the $B(\mathrm{GT})$ distributions in the $\beta$-decay of these nuclei.

We address the $N=Z$ systems from modern realistic nuclear forces derived from chiral effective field theory (EFT) using the state-of-the-art valence-space in-medium similarity renormalization group (VS-IMSRG) method. The VS-IMSRG \cite{HeikoReview2016, RagnarReview2019, RagnarPRC2016, RagnarPRL2017, MiyagiPRC2020} has emerged as a powerful \textit{ab initio} method for describing atomic nuclei from fundamental two- and three-nucleon interactions and has demonstrated considerable success in capturing key structural effects in both neutron-rich \cite{YuanN28PLB2024, YuanPRCL2024, SahooPRC2025, SahooPRCL2025, Cao2025} and proton-rich isotopes \cite{Li2023, HLi2025}, extending its applicability beyond the valley of stability. Phenomenological nuclear models typically require the introduction of an empirical quenching factor for GT transitions to reproduce experimentally observed strengths \cite{Gabriel1996, Yoshida2018, Suhonen2017}. A distinct advantage of the VS-IMSRG, together with other modern \textit{ab initio} approaches, lies in its ability to explore weak decays as well as electromagnetic observables by incorporating missing many-body correlations and consistent two-body currents (2BCs) from chiral EFT, thereby resolving the longstanding quenching puzzle \cite{Gysbers2019, Stroberg2021, Li2025, MiyagiMomentsPRL2024, Brase2026}.

This article is organized as follows. We briefly describe the VS-IMSRG framework and the details of $\beta$-decay calculations in Sec. \ref{method}, and outline the procedure for obtaining the isoscalar and isovector pairing contributions to nuclear observables. Sec. \ref{results} presents the calculated half-lives, detailed ground state properties, and $B(\mathrm{GT})$ distributions of the WP nuclei and discusses the role of pairing correlations in shaping these observables. The findings are then summarized in Sec \ref{conclusion}. %

\section{Formalism} \label{method}
In the VS-IMSRG approach\cite{RagnarReview2019}, one starts with an intrinsic A-body Hamiltonian 
\begin{equation} \label{eq1}
    H = \sum_{i}^A  (1-\frac{1}{A}) \frac{{\vec{p_i}}^2}{2m} +
         \sum_{i<j}^A \left( V_{ij}^{2N}- \frac{\vec{p_i}.\vec{p_j}}{mA} \right)+ \sum_{i<j<k}^A V_{ijk}^{3N},
\end{equation}
where $\vec{p_i}$ corresponds to nucleon momentum in the laboratory frame, $m$ denotes the nucleon mass, $V_{ij}^{2N}$ and $V_{ijk}^{3N}$ represent the two- and threee-nucleon forces respectively. The Hamiltonian is then rewritten in terms of normal-ordered operators with respect to a reference state tailored to the nucleus of interest and truncated at the two-body level \cite{RagnarPRC2016}, yielding
\begin{equation} \label{eq2}
     H=E_0+\sum_{ij}f_{ij} :a^\dagger_i a_j: + \frac{1}{4}\sum_{ijkl}\Gamma_{ijkl} :a^\dagger_i a^\dagger_j a_l a_k:.
\end{equation}
This is known as the IMSRG(2) approximation, where $E_0$, $f_{ij}$, and $\Gamma_{ijkl}$ correspond to zero-, one-, and two-body terms that capture the bulk part of $3N$ forces through ensemble normal ordering \cite{RagnarPRL2017}.

Within VS-IMSRG, an effective Hamiltonian is decoupled for a chosen valence by applying continuous unitary transformations $U(s)$ to $H$ defined in the large Hilbert space through the SRG flow equations given by
\begin{equation} \label{eq3}
\frac{dH(s)}{ds}=\left[ \eta(s),H(s)\right],
\end{equation}
where `$s$' is known as the flow parameter and $\eta(s)=\frac{dU(s)}{ds}U^{\dagger}(s)$ is called the anti-hermitian generator. In practice, the flow equations are also truncated at the two-body level, discarding induced $3N$ operators, and are solved using the Magnus formulation \cite{IMSRG_Magnus}.

In the present work, we employ the popular EM1.8/2.0 interaction \cite{EMinteraction}, generated with NuHamil \cite{NuHamil}, as the starting Hamiltonian, which provides a reliable description of g.s. energies and spectroscopic properties of nuclei over a wide mass range \cite{Miyagi2022PRC}. This interaction consists of a chiral $2N$ force at next-to-next-to-next-to-leading order (N$^3$LO) renormalized to $\lambda=1.8$ fm$^{-1}$ \cite{EM2NPotential} and a next-to-next-to-leading (N$^2$LO) order chiral $3N$ force with momentum cut-off $\Lambda=2.0$ fm$^{-1}$. 
The IMSRG calculations are performed in 13 harmonic-oscillator major shells ($e_{\textrm{max}}=12$) at $\hbar \omega=16$ MeV, with additional truncation $e_1+e_2+e_3 \leq E_{3max}=24$ on $3N$ forces to ensure convergence. 
The valence-space Hamiltonians are decoupled for the $fp$-shell, which includes the $f_{7/2}$, $f_{5/2}$, $p_{3/2}$, $p_{1/2}$ orbitals for both protons and neutrons outside a $^{40}$Ca core, utilizing the imsrg++ code \cite{IMSRGCodeRagnar}. The GT operators are also consistently evolved within the same framework, thereby eliminating the need for phenomenological quenching factors. 
Owing to normal ordering, the GT operator can be evolved with respect to either the parent or the daughter reference state. These alternative choices lead to differences of about $\sim 5\%$ in the transition matrix elements due to the truncation of many-body operators at the IMSRG(2) level \cite{Gysbers2019}. As a natural choice, we adopt the parent nucleus as the reference state for the GT operator evolution.

For $\beta$-decay, the total half-life is obtained by summing the partial decay rates from an initial state $i$ of the parent to all final states $f$ of the daughter nucleus within the kinematically allowed energy window defined by the $Q$-value,
\begin{equation}
    T_{1/2}= \left( \sum_{f} \frac{1}{t_{if}}   \right)^{-1}.
\end{equation}
The partial half-life is given by \cite{SuhonenBook}
\begin{equation}
    t_{if} = \frac{\kappa}{ f_0 B(\mathrm{GT})_{i \rightarrow f}},
\end{equation}
where $\kappa=6144.48 \pm 3.7s$ \cite{Hardy2020} and $f_0=\left( f_0^{\beta^+}+f_0^{EC} \right)$ is known as the phase space factor for $\beta^+$/EC. $f_0^{EC}$ is calculated from the expression given in Ref. \cite{Vikas2023} while $f_0^{\beta^+}$ is evaluated by solving the Fermi integral \cite{Haaranen2017}:
\begin{equation*}
    f^{\beta^+}_0 = \int_1^{W0} dW \; F(Z, W) \sqrt{W^2-1} W (W_0-W)^2,
\end{equation*}
where $W$ represents the energy of the emitted positron in units of $m_ec^2$, $W_0$ is the maximum positron energy, and $F(Z, W)$ is the generalized Fermi function that includes both Coulomb distortion of the positron wave function and finite nuclear size effects. 

The $B(\mathrm{GT})$ value between the initial and final state is defined as
\begin{equation}
    B(\mathrm{GT})_{i \rightarrow f}=\frac{g^2_A}{2J_i+1} {|\langle f || \sigma \tau||i \rangle|}^2,
\end{equation}
where $J_i$ denotes the total angular momentum of the initial state, $\sigma \tau$ is the GT operator and $g_A=1.27$ is the axial vector coupling constant. The GT strengths are computed using the Lanczos strength function method \cite{Haxton2005, Caurier2005, Utsuno2015} with sufficient Lanczos iterations to guarantee numerical convergence. The K-SHELL code is employed to perform all shell model calculations \cite{KSHELL}.

For isovector and isoscalar pairing, we follow the definitions used by Dufour and Zuker \cite{DufourZucker1996}. They demonstrated that the dominant pairing contributions to an effective interaction arise from the isovector ($L=0$, $S=0$, $J=0$, $T=1$) and isoscalar pairing ($L=0$, $S=1$, $J=1$, $T=0$) channels, and these pairing modes can be expressed by the well-known schematic operators \cite{DufourZucker1996}
\begin{equation}
P_{01} = \Omega^{-1/2} \sum_{r} \sqrt{(j_r+1/2)} Z^{\dagger}_{rr01}
\end{equation}
and
\begin{align}
    P_{10}  = & \Omega^{-1/2} \sum_{j,j'} {\left[ \frac{2(2l+1)}{(1+\delta_{jj'})} \right]}^{1/2} \times \nonumber \\
  & \{ (ll)0 \; (1/2\:1/2) 1 | (j j') 1 \} Z^{\dagger}_{rr'10} \;,
\end{align}
respectively. Here, $r \equiv (nlj)$ labels a single-particle orbit and $Z^{\dagger}_{rr'JT}$ denotes a normalized pair-creation operator coupling two nucleons to total angular momentum $J$ and isospin $T$. $\Omega = \sum_r (j_r+1/2)$ ensures proper normalization of operators in a given shell and $\{ (ll)0 \;(1/2\:1/2) 1 | (j_r j_{r'}) 1 \}$ is the coefficient for $LS$ to $jj$ transformation.

It was shown that the leading eigenmodes in the isoscalar and isovector channels, obtained from diagonalizing the pairing terms of a realistic $NN$ effective interaction, are well described by the schematic pairing operators \cite{DufourZucker1996}. A similar conclusion was reached using phenomenological shell-model interaction in Ref. \cite{PovesPLB1998}, where these operators were employed to investigate pairing contributions to nuclear observables.

\begin{table}[h]
    \centering
    \caption{Eigenvalue ($E$) and eigenvectors ($U_{rr'}$) computed from pairing Hamiltonian are compared with those of the $P_{01}$ and $P_{10}$ operators. The single-particle orbits ($rr'$) are labeled by their $2j$ values.}
    \label{tab:PairinStrength}
    \setlength{\tabcolsep}{6pt} 
    \renewcommand{\arraystretch}{1.2} 
    \begin{tabular}{lcccc}
    \hline \hline
    Orbits ($rr'$) & $P01$ & $U^{01}_{rr'}$ & $P10$ & $U^{10}_{rr'}$ \\
    \hline 
    77 & 0.63 & 0.68 & 0.41  &  0.45 \\
    75 &      &      & -0.67 & -0.69 \\
    33 & 0.45 & 0.34 & -0.27 & -0.09 \\
    35 &      &      &     0 & 0.29  \\
    31 &      &      & 0.33  & 0.28 \\
    55 & 0.55 & 0.59 & -0.42 & -0.38 \\
    11 & 0.32 & 0.25 & -0.10 &  0.07 \\
    \\ 
    \hline
    $E$ &     & -4.69 &     &  -4.49 \\
    
    \hline \hline
    \end{tabular}
\end{table}

To examine whether schematic pairing operators capture the pairing correlations in Hamiltonians derived from $2N$ and $3N$ forces within the VS-IMSRG framework, we perform an analysis analogous to that of Ref. \cite{DufourZucker1996}. Monopole-free pairing Hamiltonians are constructed and diagonalized separately in the isoscalar and isovector channels. The eigenvectors associated with the dominant eigenvalues are then compared with the corresponding schematic operators values in Table \ref{tab:PairinStrength}. The resulting pairing eigenmodes are closely reproduced by the schematic forms, and we therefore adopt these pairing definitions. The corresponding eigenvalues are then considered as the pairing strength $G$ for the schematic operators, and the explicit expressions for the two-body matrix element of these operators in $jj$-coupling are given by  \cite{PovesPLB1998}

\begin{align}
\langle j_a j_b JT | P_{01} | & j_c j_d JT  \rangle  = \nonumber \\
& \Omega^{-1} G_{IV} \sqrt{(j_a + \tfrac{1}{2})(j_c + \tfrac{1}{2})}
\, \delta_{ab}\,\delta_{cd}\,\delta_{J0}\,\delta_{T1}
\end{align}

and

\begin{align}
\langle j_a j_b JT | P_{10} | j_c j_d & JT \rangle = \nonumber \\
 & \Omega^{-1}G_{IS}
\frac{2(-1)^{j_a - j_c}}
{\sqrt{1+\delta_{ab}}\,\sqrt{1+\delta_{cd}}} \nonumber \\
& \times
\sqrt{(2j_a+1)(2j_b+1)(2j_c+1)(2j_d+1)} \nonumber \\
& \times
\begin{Bmatrix}
\tfrac{1}{2} & j_a & l_a \\
j_b & \tfrac{1}{2} & 1
\end{Bmatrix}
\begin{Bmatrix}
\tfrac{1}{2} & j_c & l_c \\
j_d & \tfrac{1}{2} & 1
\end{Bmatrix} \nonumber \\
& \times \delta_{AB}\, \delta_{CD}\, \delta_{J1}\,\delta_{T0} \:,
\end{align}
where the upper and lower case letters correspond to the $nl$ and $nlj$ quantum numbers of an orbit.

Within VS-IMSRG, the inclusion of $3N$ forces primarily modifies the monopole components of the effective interaction relative to that derived from realistic $2N$ forces alone \cite{RagnarReview2019, ZuckerPRL2003}. As a result, the pairing eigenmodes extracted from VS-IMSRG Hamiltonians are consistent with those observed in Ref. \cite{DufourZucker1996}. We find that the conclusions of Table \ref{tab:PairinStrength} remain unchanged when the same analysis is performed using VS-IMSRG interactions derived from only $2N$ forces. But the resulting pairing strengths $G$ are reduced by approximately 1 MeV. Finally, we note that within VS-IMSRG, the isovector and isoscalar pairing strengths ($G_{IV}$ and $G_{IS}$) are comparable in magnitude while they relatively differ (by $\sim$ 1.5 MeV) in the phenomenological models \cite{DufourZucker1996, PovesPLB1998}.

To understand the roles played by isoscalar and isovector pairing, we follow the strategy of Ref. \cite{PovesPLB1998}. We subtract the $P_{01}$ and $P_{10}$ pairing terms from the VS-IMSRG interactions and study the physical systems with new VS-IMSRG-$P_{01}$ and VS-IMSRG-$P_{10}$ interactions. The contribution of each pairing channel is then quantified by comparing the resulting observables with those obtained from the reference VS-IMSRG calculation.

Since the VS-IMSRG Hamiltonians are nucleus-dependent, the extracted pairing strengths $G$ vary slightly across isotopes. Table \ref{tab:PairinStrength} corresponds to the results for $^{52}$Fe nucleus. In this work, we focus on studying the pairing effect in heavier $N=Z$ isotopes in the $A \approx 50-70$ mass region. In this mass range, $G_{IV}$ varies from -4.69 to -5.18 while $G_{IS}$ exhibits a marginal change from -4.49 to -4.68. However, these variations have a rather negligible impact on the calculated observables. Therefore, for consistency, we adopt average values $G_{IV}=-4.91$ and $G_{IS}=-4.56$ for all isotopes in the present calculations.

\section{Results and Discussion} \label{results}
\begin{figure}[h]
    \centering
    \includegraphics[scale=0.70]{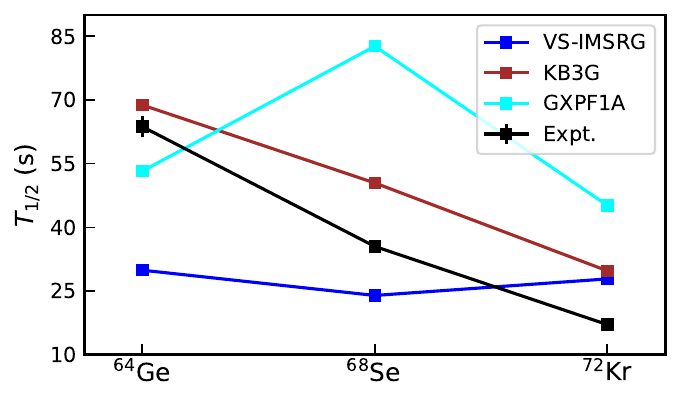}
    \caption{Half-lives of the WP nuclei.}
    \label{fig:T_half}
\end{figure}

The half-lives of the WP nuclei are evaluated from their $\beta$-decay transitions to the $1^+$ states of the daughter nucleus within the experimentally determined $Q$-value window \cite{NNDC_Q}. The resulting half-lives are compared with experimental data as well as with calculations based on 
the widely used shell model interactions KB3G \cite{kb3g} and GXPF1A \cite{gxpf1a}, employing a phenomenological quenching factor of $q=0.77$, as shown in Fig.~\ref{fig:T_half}. For $^{72}$Kr and $^{68}$Se, the half-lives obtained from VS-IMSRG are consistent with both experimental data and KB3G results, while they are overestimated by the GXPF1A interaction. But the calculated $T_{1/2}$ is found to be severely underestimated in $^{64}$Ge. The discrepancies with VS-IMSRG results can be attributed to the IMSRG(2) approximation, which may not adequately capture the full spectrum of $1^+$ states in the daughter nuclei within the defined $Q$-value, thereby affecting the predicted $T_{1/2}$. More advanced extensions of the IMSRG framework, such as various approximations incorporating IMSRG(3) effects \cite{IMSRG3f2, IMSRG3N7_Heinz2021, IMSRG3N7_Ragnar, IMSRG3N7_Heinz2024}, have demonstrated improved spectroscopic accuracy and are expected to provide a more reliable description of $T_{1/2}$.

While two different model spaces, the $fp$ and $f5pg9$ (comprising of $f_{5/2}$, $p$, and $g_{9/2}$ orbitals), can be employed to describe nuclei in this mass region, recent studies indicate that the $g_{9/2}$ orbital play a limited role in the low-energy structure and the $B(\mathrm{GT})$ distributions are adequately captured within $fp$-shell in nuclei with $A \sim 60-70$ \cite{Nichols2014, Kumar2022}. As will be discussed later, the dominant decay branches of the WP nuclei are also concentrated at low-excitation energies. We also decoupled VS-IMSRG interactions in the $f5pg9$ model space; however, the calculated half-lives are overestimated, and no significant improvement is observed. So, the present work adopts the $fp$-model space. Nevertheless, the extended model space $f5pg9$ becomes important for describing nuclear structure at higher excitation energies in nuclei with $A \sim 70$. Future measurements on $B(\mathrm{GT})$ in this regime are needed for assessing the role of $g_{9/2}$ orbital as well as the sensitivity to contributions from $f_{7/2}$ orbital. These aspects, including their role in heavier WP nuclei, require a separate study.

\begin{figure*}
    \centering
    \includegraphics[scale=0.68]{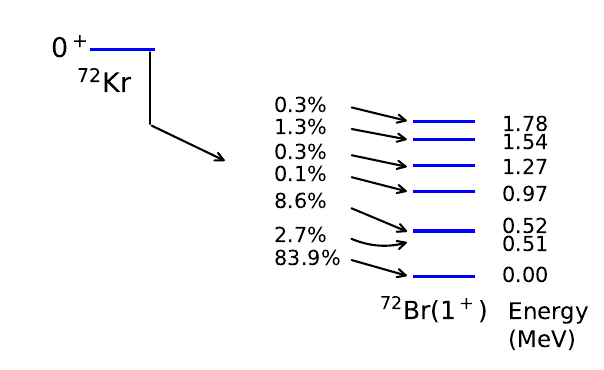}
    \includegraphics[scale=0.68]{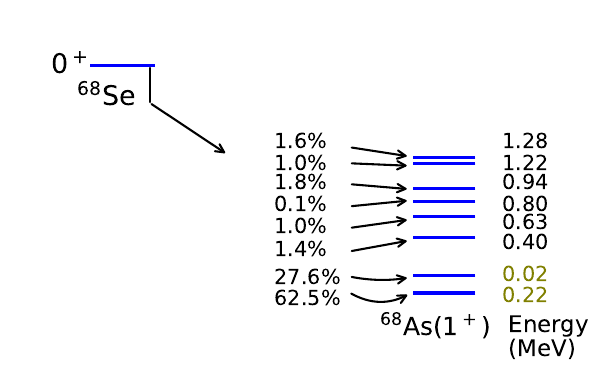}
    \includegraphics[scale=0.68]{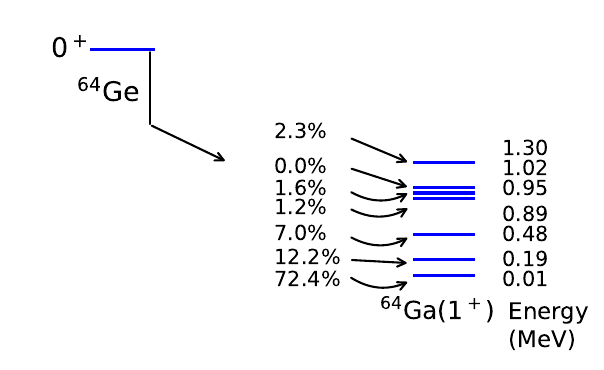}
    \caption{$\beta^+$-decay scheme of $^{72}$Kr, $^{68}$Se, and $^{64}$Ge illustrating the dominant branching ratios to different $1^+$ states of the daughter nuclei. The energies shown in olive correspond to states appearing below the g.s. of daughter nuclei.}
    \label{fig:BranchingRatios}
\end{figure*}

Fig.~\ref{fig:BranchingRatios} illustrates the decay schemes of the WP nuclei, highlighting the dominant decay branches to daughter states relative to the observed g.s., in terms of branching ratios (B.R.). For the $k^{th}$ state, the B.R. is computed as (B.R.)$^k=t^k_{if}/T_{1/2}$, representing the fraction of total decay rate that feeds the $k^{th}$ final state. In the absence of comprehensive experimental data, these predicted branching patterns may provide useful guidance for identifying prominent decay branches in future measurements and offer a reliable reference for benchmarking the total decay distribution. For $^{72}$Kr, the ground state of $^{72}$Br constitutes the strongest decay branch, followed by the the $1^+$ states at 0.5 MeV. 
 The decay of $^{68}$Se is almost concentrated at the first two $1^+$ states of $^{68}$As, which together account for 90$\%$ of the total branching. In the VS-IMSRG calculations, however, these states appear below the g.s. of $^{68}$As, in contrast to experimental observation.
A similar pattern is observed in the decay of $^{64}$Ge, where the first three $1^+$ states contribute 91.6$\%$ of the total decay branching. Overall, nearly 95$\%$ of the total decay intensity of these WP nuclei is confined within an excitation energy range of $\approx$ 0.6 MeV.

While a more accurate description of $T_{1/2}$ can be achieved by precise calculations of excitation energies of the daughter states and refined phase-space factors \cite{Stoica2016}, the dominant role is played by the underlying GT strength distributions. Recent \textit{ab initio} studies have shown that implementing GT operators with consistent 2BCs from chiral EFT can lead to improved agreement with experimental data \cite{Gysbers2019, Stroberg2021, Li2025}. In the present work, however, we do not include 2BCs; instead, we concentrate on elucidating the impact of isoscalar and isovector pairing correlations on the $B(\mathrm{GT})$ distributions. We first analyze their contributions to the ground-state structure of the WP nuclei and then examine how these correlations influence the energy distribution of $B(\mathrm{GT})$ strengths.

The ground states of even-even nuclei are commonly understood in terms of condensate-like many-body states built from coherent nucleon pairs. While they are typically interpreted in terms of isovector ($L=0$, $S=0$, $T=1$) pairing condensate corresponding to seniority-zero configurations, there have been conjectures for the possible existence of isoscalar ($L=0$, $S=1$, $T=0$) pair condensate in heavier $N=Z$ nuclei. To examine this, we investigate the probability of finding the physical ground states of $N=Z$ systems in ideal pairing condensates by computing ${\langle \Psi_{g.s}|\Psi_{P_{01}} \rangle}^2$ and ${\langle \Psi_{g.s}|\Psi_{P_{10}} \rangle}^2$, the square of overlaps of the g.s. with isovector and isoscalar condensate states. The isovector condensate is obtained by diagonalizing the $P_{01}$ Hamiltonian in the $fp$-shell using VS-IMSRG single-particle energies. The isoscalar condensate is computed from the $P_{10}$ interaction using VS-IMSRG single-particle energies with the spin–orbit splitting removed, as isoscalar pairs are formed within degenerate $l$-multiplets. The calculated overlaps reveal no traces of the isoscalar condensate in the ground states, $\langle \Psi_{g.s}|\Psi_{P_{10}} \rangle$ being smaller than $1\%$ across all $N=Z$ isotopes. The overlaps with the isovector condensates are summarized in Table \ref{tab:overlap}.

\begin{table}[h]
    \centering
    \caption{Overlaps between physical ground state and isovector pairing condensates in $N=Z$ nuclei}
    \label{tab:overlap}
    \setlength{\tabcolsep}{6pt} 
    \renewcommand{\arraystretch}{1.2} 
    \begin{tabular}{lcccccccccc}
    \hline \hline
    Nucleus \hspace{3mm}& ${\langle \textrm{VS-IMSRG}|P_{01} \rangle}^2$ &\hspace{3mm}${\langle \textrm{KB3G}|P_{01} \rangle}^2$ \\
    \hline
        $^{72}$Kr & 10.1 &2.1\\
        $^{68}$Se & 4.3  &0.2\\
        $^{64}$Ge & 6.8  &1.3\\
        $^{60}$Zn & 15.0 &32.8\\
        $^{56}$Ni & 48.7 &72.9\\
        $^{52}$Fe & 30.5 &42.3\\
    
         \hline \hline
    \end{tabular}
\end{table}

As shown in Table \ref{tab:overlap}, the ground states of WP nuclei have a marginal overlap with the isovector condensates. We also compared the VS-IMSRG results with those obtained from KB3G interaction, with pairing strengths and operators adopted for its original version KB3 \cite{PovesPLB1998}. Both interactions lead to consistent conclusions, confirming the absence of a pronounced isovector condensate in the WP nuclei. For other $N=Z$ systems, the overlaps are also small, though not marginal, and therefore do not support a genuine isovector-condensate picture. The ground states of $^{56}$Ni and $^{52}$Fe are dominated by $(0f_{7/2})^n$ configurations, where the proton-neutron interactions are weak, and the like-particles are paired to $J=0$. Thus, they have a reasonable overlap with the isovector condensate. Being doubly magic, $^{56}$Ni has the largest overlap and a nearly $50\%$ probability of being found in the isovector condensate. The overlaps obtained for $^{52}$Fe, $^{56}$Ni, and $^{60}$Zn with KB3G are systematically larger than those calculated with VS-IMSRG.

\begin{figure}[h]
    \centering
    \includegraphics[scale=0.68]{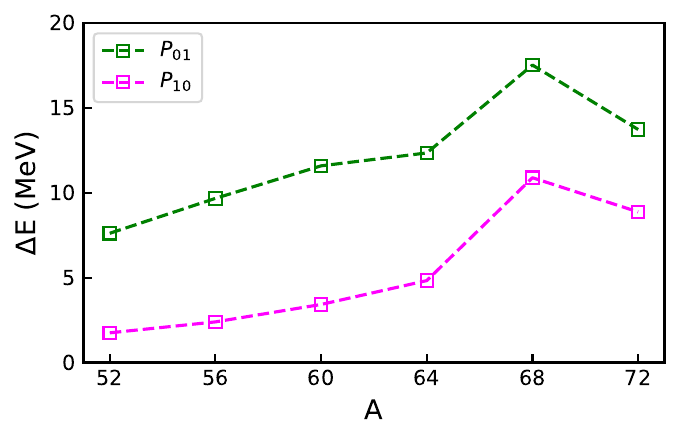}
    \caption{Pairing contributions to the g.s energies of $N=Z$ nuclei.}
    \label{fig:gs_energy}
\end{figure}

Fig. \ref{fig:gs_energy} depicts the pairing contributions to g.s. energies in valence space for $N=Z$ nuclei. They are obtained by subtracting the g.s. energies of VS-IMSRG-$P_{01}$ and VS-IMSRG-$P_{10}$ Hamiltonians from the VS-IMSRG results. The pairing contributions exhibit similar patterns in both channels, gradually increasing with mass number from $A=52$ to 64, then steeply rising at $A=68$. The isovector contributions to g.s. energies remain larger than the isoscalar ones across all $N=Z$ isotopes.

\begin{figure*}
    \centering
    \includegraphics[scale=0.65]{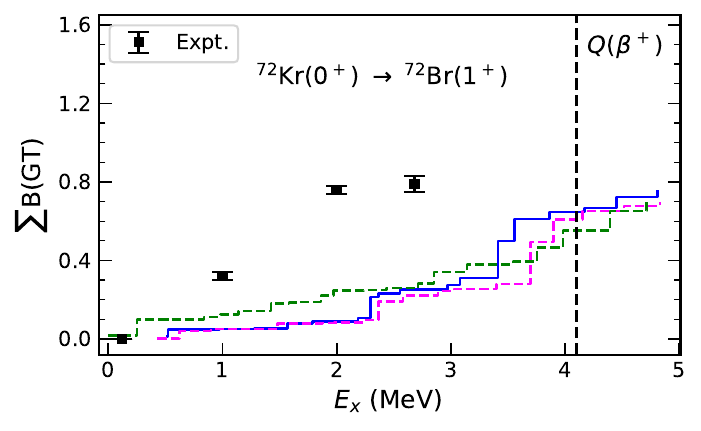}
    \includegraphics[scale=0.65]{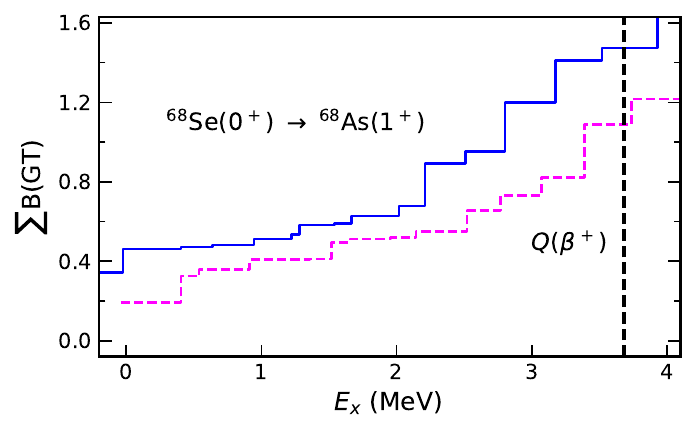}
    \includegraphics[scale=0.65]{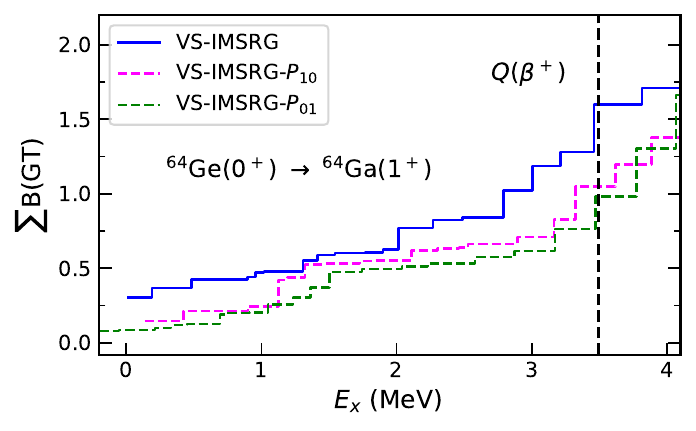}
    \caption{Cumulative GT strength for the $\beta^+$-decay of WP nuclei within the $Q$-value energy window \cite{NNDC_Q} as a function of excitation energy of the final nucleus.}
    \label{fig:BGT_Sum}
\end{figure*}

\begin{figure*}
    \centering
    \includegraphics[scale=0.65]{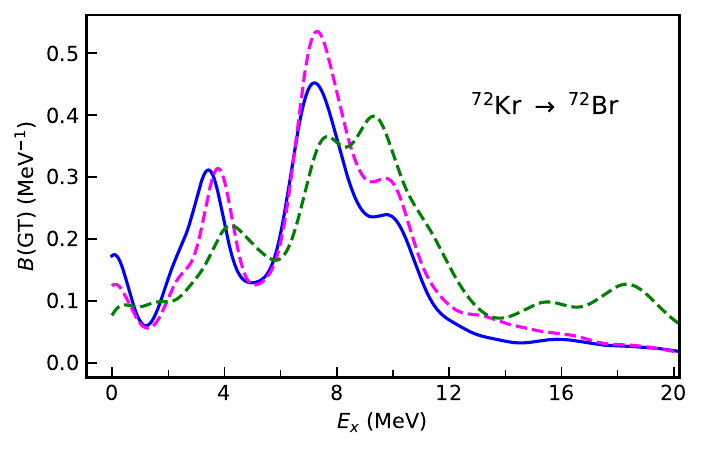}
    \includegraphics[scale=0.65]{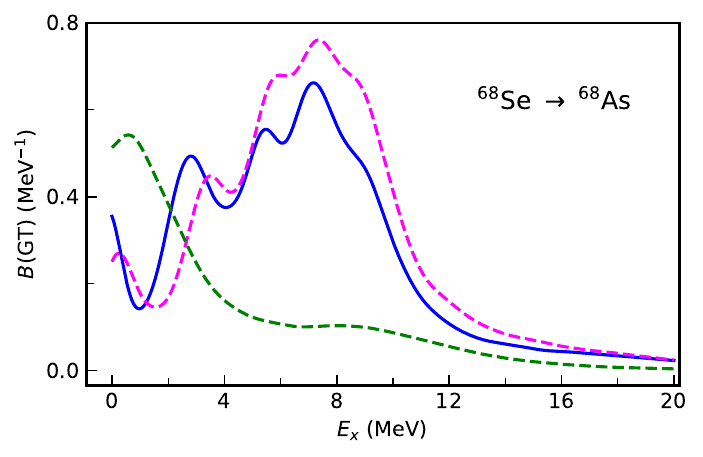}
    \includegraphics[scale=0.65]{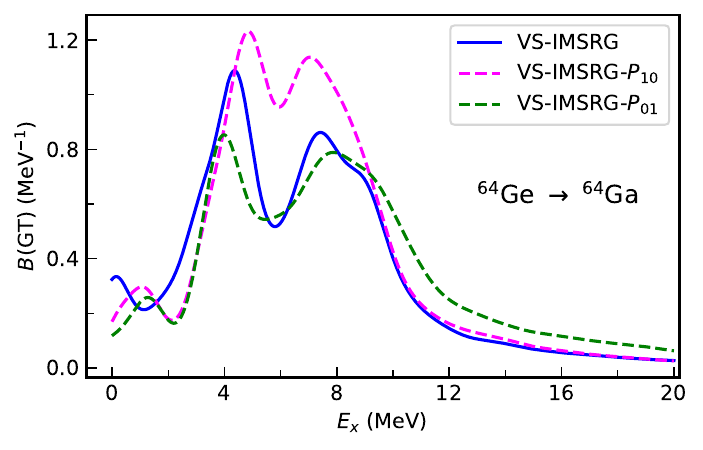}
    \caption{ $B(\mathrm{GT})$ strength distribution in $^{72}$Br, $^{68}$As, and $^{64}$Ga, as a function of excitation energy.}
    \label{fig:BGT_distributionWP}
\end{figure*}

The isovector and isoscalar pairing contributions to the sum of $B(\mathrm{GT})$ values within the low-energy $Q$-value window in WP nuclei are shown in Fig. \ref{fig:BGT_Sum}. Their contributions can be realized by comparing the accumulated $B(\mathrm{GT})$ values over excitation energy ($E_x$) obtained from VS-IMSRG-$P_{01}$ and VS-IMSRG-$P_{10}$ with those of the original VS-IMSRG calculations. Deviations from the latter thus reflect the relative impact of the corresponding pairing correlations on the cumulative $B(\mathrm{GT})$ distribution. As shown in Fig. \ref{fig:gs_energy}, the isovector pairing terms contribute significantly to the g.s. energies of WP nuclei, particularly for $^{68}$Se. A similarly strong effect is observed for their $\beta^+$-decay daughter nuclei, where the isovector contribution to the g.s. energy can even exceed that of several low-lying $1^+$ excited states. For example, within VS-IMSRG-$P_{01}$, many $1^+$ states appear below the observed g.s. of $^{68}$As, thereby preventing a direct one-to-one comparison of the $B(\mathrm{GT})$ distributions with the original VS-IMSRG results. For this reason, the cumulative $B(\mathrm{GT})$ obtained from VS-IMSRG-$P_{01}$ for $^{68}$As is not shown in Fig.~\ref{fig:BGT_Sum}. In contrast, for the other daughter nuclei, either no or at most one or two $1^+$ states lie below the observed ground state, allowing for a meaningful comparison.

Experimentally, an accumulated strength of $0.79(4)$ is reported up to $E_x=2.7$ MeV in $^{72}$Br, whereas the calculations reproduce a similar value of around $0.7$ at a somewhat higher excitation energy of $\sim 4$ MeV. This difference may arise from the IMSRG(2) truncation, which tends to shift excited states to higher energies. 
In the low-energy region, the isoscalar pairing terms have a negligible impact, and thus the accumulated $B(\mathrm{GT})$ values in $^{72}$Br remain almost the same as the original VS-IMSRG results when isoscalar pairing correlations are not included. The isovector pairing terms also have a smaller impact on the $B(\mathrm{GT})$ sum. In contrast, for $^{68}$Se decay, the cumulative $B(\mathrm{GT})$ strength includes contributions from the isoscalar pairing terms, and at the corresponding $Q$-value, the accumulated $B(\mathrm{GT})$ is reduced by approximately 0.3 units when these contributions are omitted.
In the case of $^{64}$Ga, both isoscalar and isovector terms make similar contributions to the accumulated $B(\mathrm{GT})$ within the $Q$-value window and at $Q(\beta^+)=3.495$ MeV, the cumulating $B(\mathrm{GT})$ strength is approximately 0.6 units smaller than actual calculated value when either isoscalar or isovector pairing correlations are excluded. Overall, in the $Q$-value window of $^{72}$Kr and $^{64}$Ge decay, both isoscalar and isovector pairing have nearly identical contributions that can modify the accumulated $B(\mathrm{GT})$ strength by $\approx 20-40\%$.

We have shown that the impact of pairing correlations on the accumulated $B(\mathrm{GT})$ strengths in the low-energy region in the $\beta^+$-decay of WP nuclei. It is therefore instructive to examine the complete $B(\mathrm{GT})$ strength distributions and how they are influenced by isovector and isoscalar pairing correlations over the entire excitation-energy range. We present all the GT strengths in the daughter nuclei $^{72}$Br, $^{68}$As, and $^{64}$Ga and illustrate the overall impact of pairing correlations on them in Fig. \ref{fig:BGT_distributionWP}. The discrete $B(\mathrm{GT})$ values are smoothed out using a normalized Gaussian function with a width of $\sigma=0.51$ MeV.

In the $^{72}\textrm{Kr} \rightarrow ^{72}\textrm{Br}$ decay, two prominent GT peaks are observed at excitation energies of around 3.5 and 7.5 MeV. A nearly similar pattern is obtained with the VS-IMSRG-$P_{10}$ Hamiltonian. But, within VS-IMSRG-$P_{01}$ interaction, these peaks are slightly shifted (by $\sim 0.5$ MeV) and reduced in strength. At $E_x \gtrsim 10$ MeV, the $B(\mathrm{GT})$ distribution from VS-IMSRG-$P_{01}$ suggests that isovector pairing correlations suppress $B(\mathrm{GT})$ strengths at high excitation energies. As discussed earlier, the exclusion of isovector pairing leads to several $1^+$ states appearing below the observed g.s. in $^{68}$As, where a significant portion of the $B(\mathrm{GT})$ strengths are concentrated. Apart from this shift, the overall $B(\mathrm{GT})$ distribution of $^{68}$As remains largely unchanged, with small variations in the 4–10 MeV region when isoscalar pairing is omitted. In the case of $^{64}\textrm{Ge} \rightarrow ^{64}\textrm{Ga}$, two GT peaks are identified at excitation energies around $\approx$4 and 8 MeV, respectively. The isovector pairing enhances the strength of the first peak and suppresses the $B(\mathrm{GT})$ strengths at higher excitation energies, where its overall influence remains weak. The isoscalar pairing also leads to a reduction of the $B(\mathrm{GT})$ strength in the 4–8 MeV energy range, with negligible impact at higher excitation energies ($E_x \gtrsim10$ MeV).

\section{Summary} \label{conclusion}
In this work, we study the $\beta$-decay properties of WP nuclei $^{72}$Kr, $^{68}$Se, and $^{64}$Ge, relevant to $rp$-process nucleosynthesis, starting from realistic two- and three-nucleon forces using the VS-IMSRG method.
The calculated branching ratios show that the dominant decay branches are confined below 1 MeV of excitation energy in the daughter nuclei. The computed half-lives indicate the need for a higher-order many-body approximation, as well as for consistent 2BCs within the \textit{ab initio} framework, to provide a more reliable description of half-lives in proton-rich systems. We investigate isoscalar and isovector pairing in the WP nuclei, together with several other $N=Z$ isotopes in the $fp$-shell region. Our results indicate no traces of isoscalar condensate and do not support any strong isovector pairing condensate in their ground states. Despite the comparable magnitudes of the extracted isoscalar and isovector pairing strengths, the isovector pairing contributions to ground state energies remain systematically larger. We observed that in the low-energy $Q$-value window, the accumulated $B(\mathrm{GT})$ strengths are enhanced by the pairing correlations. We showed the complete $B(\mathrm{GT})$ strength distributions over excitation energies of the daughter nuclei and the overall impact of isovector and isoscalar pairing correlations on them. This work offers a microscopic understanding of $\beta$-decay properties and pairing correlations in $N=Z$ nuclei approaching the proton drip line.

\vspace{-0.5cm}

\section*{Acknowledgment}
This work is supported by the research Grant No. CRG/2022/005167 from SERB (India).
We gratefully acknowledge Dr. Takayuki Miyagi and Dr. Anil Kumar for helpful discussions regarding the implementation of the Lanczos strength function method.  We would like to thank the National Supercomputing Mission (NSM) for providing computing resources of ‘PARAM Ganga’ at the Indian Institute of Technology Roorkee, implemented by C-DAC and supported by the Ministry of Electronics and Information Technology (MeitY) and Department of Science and Technology (DST), Government of India.
\vspace{-0.5cm}
\section*{Data availability}
The data supporting this study’s findings are available within the article.


\begin{thebibliography}{44}

\bibitem{LangankeReview}
K. Langanke and G. Mart{\'i}nez-Pinedo, ``Nuclear weak-interaction processes in stars",
\href{https://doi.org/10.1103/RevModPhys.75.819}
{Rev. Mod. Phys. {\bf 75}, 819 (2003).}

\bibitem{SuzukiReview}
T. Suzuki and N. Shimizu, ``Shell-model study of weak-decays relevant to astrophysical processes",
\href{https://doi.org/10.3389/fphy.2024.1434598}
{Front. Phys. {\bf 12}, 1434598 (2024).}

\bibitem{SchatzRep}
H. Schatz, A. Aprahamian, J. Görres, M. Wiescher, T. Rauscher, J. Rembges, F.-K. Thielemann, B. Pfeiﬀer, P. Möller, K.-L. Kratz, H. Herndl, B. Brown, H. Rebel, ``$rp$-process nucleosynthesis at extreme temperature and density conditions",
\href{https://doi.org/10.1016/S0370-1573(97)00048-3}
{Phys. Rep. {\bf 294}, 167 (1998).}

\bibitem{Schatz2001}
H. Schatz, A. Aprahamian, V. Barnard, L. Bildsten, A. Cumming, M. Ouellette, T.
Rauscher, F.-K. Thielemann, M. Wiescher, ``End point of the $rp$ process on accreting neutron stars",
\href{https://doi.org/10.1103/PhysRevLett.86.3471}
{Phys. Rev. Lett. {\bf 86}, 3471 (2001).}

\bibitem{Wormer1994}
L. van Wormer, J. Goerres, C. Iliadis, M. Wiescher, F.-K. Thielemann, ``Reaction rates and reaction sequences in the $rp$-process",
\href{https://ui.adsabs.harvard.edu/link_gateway/1994ApJ...432..326V/doi:10.1086/174572}
{Astrophys. J. {\bf 432}, 326 (1994).}

\bibitem{Rubio2017}
B. Rubio, L. Kucuk, S. E. A. Orrigo, Y. Fujita, W. Gelletly, B. Blank, T. Adachi, P. Aguilera, J. Agramunt, A. Algora {\it et al.}, ``Beta decay studies of proton rich nuclei, an important ingredient for $rp$-process calculations",
\href{https://doi.org/10.7566/JPSCP.14.010510}
{JPS Conf. Proc. {\bf 14}, 010510 (2017).}


\bibitem{Parikh2013}
A. Parikh, J. José, G. Sala, C. Iliadis, ``Nucleosynthesis in type I X-ray bursts",
\href{https://doi.org/10.1016/j.ppnp.2012.11.002}
{Prog. Part. Nucl. Phys. {\bf 69}, 225 (2013).}

\bibitem{Nacher2023}
E. Nácher, S. Parra, J. A. Briz, P. Aguilera, J. Agramunt, A. Algora, T. Berry, M. J. G. Borge, M. Carmona, L. M. Fraile {\it et al.}, ``Beta decay along the $N=Z$ line and its relevance in rp-process and X-ray bursts",
\href{https://doi.org/10.1051/epjconf/202327912004}
{EPJ Web Conf. {\bf 279}, 12004 (2023).}

\bibitem{Zhou2023}
X. Zhou, M. Wang, Y.H. Zhang, Y.A. Litvinov, Z. Meisel, K. Blaum, X.H. Zhou, S.Q.
Hou, K.A. Li, H.S. Xu {\it et al.}, ``Mass measurements show slowdown of rapid proton capture process at waiting-point nucleus $^{64}$Ge",
\href{https://doi .org /10 .1038 /s41567 -023 -02034 -2}
{Nat. Phys. {\bf 19}, 1091 (2023).}

\bibitem{Chen2024}
Z.-R. Chen, L.-J. Wang, ``Stellar weak-interaction rates for $rp$-process waiting-point nuclei from projected shell model",
\href{https://doi.org/10.1016/j.physletb.2023.138338}
{Phys. Lett. B {\bf 848}, 138338 (2024).}

\bibitem{Hamamoto1995}
I. Hamamoto and X.Z. Zhang, ``Dependence of Gamow-Teller $\beta^+$-decay of $^{80}$Zr, $^{76}$Sr and $^{72}$Kr on nuclear shape",
\href{https://doi.org/10.1007/BF01295892}
{Z. Phys. A {\bf 353}, 145 (1995).}

\bibitem{Poirier2004}
E. Poirier, F. Maréchal, Ph. Dessagne, A. Algora, M. J. G. Borge, D. Cano-Ott, J. C. Caspar, S. Courtin, J. Devin, L. M. Fraile \textit{et al.}, ``$B(\mathrm{GT})$ strength from $\beta$-decay measurements and inferred shape mixing in $^{74}$Kr",
\href{https://doi.org/10.1103/PhysRevC.69.034307}
{Phys. Rev. C {\bf 69}, 034307 (2004).}

\bibitem{Nacher2004}
E. Nácher, A. Algora, B. Rubio, J. L. Ta{\'i}n, D. Cano-Ott, S. Courtin, Ph. Dessagne, F. Maréchal, Ch. Miehé, and E. Poirier \textit{et al.}, ``Deformation of the $N=Z$ nucleus $^{76}$Sr using $\beta$-decay studies",
\href{https://doi.org/10.1103/PhysRevLett.92.232501}
{Phys. Rev. Lett. {\bf 92}, 232501 (2004).}


\bibitem{Cerdan2009}
A. B. Pérez-Cerdán, B. Rubio, W. Gelletly, A. Algora, J. Agramunt, E. Nácher, J. L. Ta{\'i}, P. Sarriguren, L. M. Fraile, M. J. G. Borge \textit{et al.}, ``Deformation of Sr and Rb isotopes close to the $N = Z$ line via $\beta$-decay studies using the total absorption technique",
\href{https://doi.org/10.1103/PhysRevC.88.014324}
{Phys. Rev. C {\bf 88}, 014324 (2013).}

\bibitem{Hoff2020}
D.E.M. Hoff, A.M. Rogers, S.M. Wang, P.C. Bender, K. Brandenburg, K. Childers, J.A. Clark, A.C. Dombos, E.R. Doucet, S. Jin \textit{et al.}, ``Mirror-symmetry violation in bound nuclear ground states",
\href{https://doi.org/10.1038/s41586-020-2123-1}
{Nature {\bf 580}, 52 (2020).}

\bibitem{Algora2025}
A. Algora, A. Vitéz-Sveiczer, A. Poves, G. G. Kiss, B. Rubio, G. de Angelis, F. Recchia, S. Nishimura, T. Rodríguez, P. Sarriguren \textit{et al.}, ``Isospin Symmetry Breaking in the $^{71}$Kr and $^{71}$Br Mirror System",
\href{https://doi.org/10.1103/PhysRevLett.134.162502}
{Phys. Rev. Lett. {\bf 134}, 162502 (2025).}



\bibitem{Waniganeththi2022}
S. Waniganeththi, D.E.M. Hoff, A.M. Rogers, C.J. Lister, P.C. Bender, K. Brandenburg, K. Childers, J.A. Clark, A.C. Dombos, E.R. Doucet \textit{et al.}, ``Establishing the ground-state spin of $^{71}$Kr",
\href{https://doi.org/10.1103/PhysRevC.106.044317}
{Phys. Rev. C {\bf 106}, 044317 (2022).}


\bibitem{Wimmer2021}
K. Wimmer, W. Korten, P. Doornenbal, T. Arici, P. Aguilera, A. Algora, T. Ando, H. Baba, B. Blank, A. Boso \textit{et al.}, ``Shape Changes in the Mirror Nuclei $^{70}$Kr and $^{70}$Se",
\href{https://doi.org/10.1103/PhysRevLett.126.072501}
{Phys. Rev. Lett. {\bf 126}, 072501 (2021).}

\bibitem{Lenzi2021}
S.M. Lenzi, A. Poves, A.O. Macchiavelli, ``Shell model analysis of the $B(E2; 2^+ \rightarrow 0^+)$ values in the $A = 70$, $T = 1$ triplet $^{70}$Kr, $^{70}$Br, and $^{70}$Se",
\href{https://doi.org/10.1103/PhysRevC.104.L031306}
{Phys. Rev. C {\bf 104}, L031306 (2021).}

\bibitem{Sveiczer2022}
A. Vitéz-Sveiczer, A. Algora, A.I. Morales, B. Rubio, G.G. Kiss, P. Sarriguren, P. Van Isacker, G. de Angelis, F. Recchia, S. Nishimura \textit{et al.}, ``The $\beta$-decay of $^{70}$Kr to $^{70}$Br: Restoration of the pseudo-SU(4) symmetry",
\href{https://doi.org/10.1016/j.physletb.2022.137123}
{Phys. Lett. B {\bf 830}, 137123 (2022).}


\bibitem{Sarriguren2009}
P. Sarriguren, ``Weak interaction rates for Kr and Sr waiting-point nuclei under $rp$-process conditions",
\href{https://doi.org/10.1016/j.physletb.2009.09.046}
{Phys. Lett. B {\bf 680}, 438 (2009).}

\bibitem{Sarriguren2011}
P. Sarriguren, ``Stellar weak decay rates in neutron-deficient medium-mass nuclei",
\href{http://dx.doi.org/10.1103/PhysRevC.83.025801}
{PHYSICAL REVIEW C {\bf 83}, 025801 (2011.}

\bibitem{Jameel2012}
J.-U. Nabi, ``rp-process weak-interaction mediated rates of waiting-point nuclei",
\href{https://doi.org/10.1007/s10509-012-0995-8}
{Astrophys. Space Sci. {\bf 339}, 305 (2012).}

\bibitem{Langanke2000}
K. Langanke, G. Martínez-Pinedo,, ``Shell-model calculations of stellar weak interaction rates: II. Weak rates for nuclei in the mass range $A=45-65$ in supernovae environments",
\href{https://doi.org/10.1016/S0375-9474(00)00131-7}
{Nucl. Phys. A {\bf 673}, 481 (2000).}

\bibitem{Langanke2001}
K. Langanke, G. Martínez-Pinedo,, ``Rate tables for the weak processes of $pf$-shell nuclei in stellar envirnoments",
\href{https://doi.org/10.1006/adnd.2001.0865}
{At. Data and Nucl. Data Tables {\bf 79}, 1 (2001).}

\bibitem{Briz2015}
J. A. Briz, E. Nácher, M. J.G. Borge, A. Algora, B. Rubio, Ph. Dessagne, A. Maira, D. Cano-Ott, S. Courtin, D. Escrig \textit{et al.}, ``Shape study of the $N = Z$ nucleus $^{72}$Kr via $\beta$-decay",
\href{https://doi.org/10.1103/PhysRevC.92.054326}
{Phys. Rev. C {\bf 92}, 054326 (2015).}


\bibitem{PovesPLB1998}
A. Poves, G. Martinez-Pinedo, ``Pairing and the structure of the $pf$-shell $N\sim Z$ nuclei",
\href{https://doi.org/10.1016/S0370-2693(98)00538-3}
{Phys. Lett. B {\bf 430}, 203 (1998).}

\bibitem{Brown2013}
B.A. Brown, ``Pairing and shell gaps in nuclei",
\href{https://doi.org/10.1088/1742-6596/445/1/012010}
{J. Phys.: Conf. Ser. {\bf 445}, 012010 (2013).}

\bibitem{Dean2003}
D.J. Dean and M. Hjorth-Jensen, ``Pairing in nuclear systems: from neutron stars to finite nuclei",
\href{https://doi.org/10.1103/RevModPhys.75.607}
{Rev. Mod. Phys. {\bf 75}, 607 (2003).}

\bibitem{Tsunoda2020}
N. Tsunoda, T. Otsuka, K. Takayanagi, N. Shimizu, T. Suzuki, Y. Utsuno, S. Yoshida, and H. Ueno, ``The impact of nuclear shape on the emergence of the neutron dripline",
\href{https://doi.org/10.1038/s41586-020-2848-x}
{Nature {\bf 587}, 66 (2020).}

\bibitem{Afanasjev2005}
A.V. Afanasjev and S. Frauendorf, ``Description of rotating $N = Z$ nuclei in terms of isovector pairing",
\href{https://doi.org/10.1103/PhysRevC.71.064318}
{Phys. Rev. C {\bf 71}, 064318 (2005).}

\bibitem{Frauendorf2014}
S. Frauendorf, A.O. Macchiavelli, ``Overview of neutron–proton pairing",
\href{http://dx.doi.org/10.1016/j.ppnp.2014.07.001}
{Prog. Part. Nucl. Phys. {\bf 78}, 24 (2014).}

\bibitem{Macchiavelli2000}
A.O. Macchiavelli, P. Fallon, R.M. Clark, M. Cromaz, M.A. Deleplanque, R.M. Diamond, G.J. Lane, I.Y. Lee, F.S. Stephens, C.E. Svensson, K. Vetter, and D. Ward, ``Is there $np$ pairing in $N = Z$ nuclei?",
\href{ https://doi.org/10.1103/PhysRevC.61.041303}
{Phys. Rev. C {\bf 61}, 041303(R) (2000).}

\bibitem{Sagawa2013}
H. Sagawa, Y. Tanimura, and K. Hagino, ``Competition between $T = 1$ and $T = 0$ pairing in $pf$ -shell nuclei with $N=Z$",
\href{https://doi.org/10.1103/PhysRevC.87.034310}
{Phys. Rev. C {\bf 87}, 034310 (2013).}


\bibitem{Cederwall2011}
B. Cederwall, F. Ghazi Moradi, T. Bäck, A. Johnson, J. Blomqvist, E. Clément, G. de France, R. Wadsworth, K. Andgren, K. Lagergren \textit{et al.}, ``Evidence for a spin-aligned neutron–proton paired phase from the level structure of $^{92}$Pd",
\href{https://doi.org/10.1038/nature09644}
{Nature {\bf 469}, 68 (2011).}

\bibitem{Engel1997}
J. Engel, S. Pittel, M. Stoitsov, P. Vogel, and J. Dukelsky, ``Neutron-proton correlations in an exactly solvable model",
\href{https://doi.org/10.1103/PhysRevC.55.1781}
{Phys. Rev. C {\bf 55}, 1781 (1997).}

\bibitem{Kaneko2018}
K. Kaneko, Y. Sun, and T. Mizusaki, ``Isoscalar neutron-proton pairing and SU(4)-symmetry breaking in Gamow-Teller transitions",
\href{https://doi.org/10.1103/PhysRevC.97.054326}
{Phys. Rev. C {\bf 97}, 054326 (2018).}

\bibitem{Bai2013}
C.L. Bai, H. Sagawa, M. Sasano, T. Uesaka, K. Hagino, H. Q. Zhang, X. Z. Zhang, F.R. Xu, ``Role of $T = 0$ pairing in Gamow–Teller states in $N = Z$ nuclei",
\href{http://dx.doi.org/10.1016/j.physletb.2012.12.060}
{Phys. Lett. B {\bf 719}, 116 (2013).}

\bibitem{Choudhary2025}
P. Choudhary and C. Qi, ``$\beta$-decay properties of $N = Z$ nuclei: Role of neutron-proton pairing and the shell model interpretation",
\href{https://doi.org/10.1103/PhysRevC.111.034316}
{Phys. Rev. C {\bf 111}, 034316 (2025).}




\bibitem{HeikoReview2016}
H. Hergert, S.K. Bogner, T.D. Morris, A. Schwenk, K. Tsukiyama, ``The In-Medium Similarity Renormalization Group: A Novel \textit{Ab Initio} Method for Nuclei",
\href{https://doi.org/10.1016/j.physrep.2015.12.007}
{Phys. Reports {\bf 621}, 165 (2016).}

\bibitem{RagnarReview2019}
S.R. Stroberg, H. Hergert, S.K. Bogner, and J.D. Holt, ``Nonempirical Interactions for the Nuclear Shell Model: An Update",
\href{https://doi.org/10.1146/annurev-nucl-101917-021120}
{Annu. Rev. Nucl. Part. Sci. {\bf 69}, 307 (2019).}

\bibitem{RagnarPRC2016}
S.R. Stroberg, H. Hergert, J.D. Holt, S.K. Bogner, and A. Schwenk, ``Ground and excited states of doubly open-shell nuclei from {\it ab initio} valence-space Hamiltonians",
\href{http://dx.doi.org/10.1103/PhysRevC.93.051301}
{Phys. Rev. C {\bf 93}, 051301(R) (2016).}

\bibitem{RagnarPRL2017}
S.R. Stroberg, A. Calci, H. Hergert, J.D. Holt, S.K. Bogner, R. Roth and A. Schwenk, ``Nucleus-Dependent Valence-Space Approach to Nuclear Structure",
\href{http://dx.doi.org/10.1103/PhysRevLett.118.032502}
{Phys. Rev. Lett. {\bf 118}, 032502 (2017).}

\bibitem{MiyagiPRC2020}
T. Miyagi, S.R. Stroberg, J.D. Holt and N. Shimizu, ``{\it Ab initio} multishell valence-space Hamiltonians and the island of inversion",
\href{http://dx.doi.org/10.1103/PhysRevC.102.034320}
{Phys. Rev. C {\bf 102}, 034320 (2020).}

\bibitem{YuanN28PLB2024}
Q. Yuan, J.G. Li, and H.H. Li, ``\textit{Ab initio} calculations for well deformed nuclei: $^{40}$Mg and $^{42}$Si",
\href{https://doi.org/10.1016/j.physletb.2023.138331}
{Phys. Lett. B {\bf 848}, 138331 (2024).}

\bibitem{YuanPRCL2024}
Q. Yuan, J.G. Li, and W. Zuo, ``\textit{Ab initio} calculations for configuration-coexisting states in $^{45}$S: An extension from $^{43}$S", 
\href{https://doi.org/10.1103/PhysRevC.109.L041301}
{Phys. Rev. C {\bf 109}, L041301 (2024).}

\bibitem{SahooPRC2025}
S. Sahoo and P.C. Srivastava, ``{\it Ab initio} study of the island of inversion in odd-A nuclei: Structure of $^{31,33}$Mg",
\href{https://doi.org/10.1103/PhysRevC.111.054308}
{Phys. Rev. C {\bf 111}, 054308 (2025).}

\bibitem{SahooPRCL2025}
S. Sahoo and P.C. Srivastava, ``Evolution of shell structure at $N=32$ and 34: Insights from realistic nuclear forces",
\href{https://doi.org/10.1103/423y-znv8}
{Phys. Rev. C {\bf 112}, L021301 (2025).}

\bibitem{Cao2025}
X.C. Cao, C.F. Jiao, ``{\it Ab initio} study in the island of inversion within the two-major-shell valence space",
\href{https://doi.org/10.1016/j.physletb.2025.140034}
{Phys. Lett. B {\bf 871}, 140034 (2025).}



\bibitem{Li2023}
J.G. Li, H.H. Li, S. Zhang, Y.M. Xing, and W. Zuo, ``Double-magicity of proton drip-line nucleus $^{22}$Si with ab initio calculation",
\href{https://doi.org/10.1016/j.physletb.2023.138197}
{Phys. Lett. B {\bf 846}, 138197 (2023).}

\bibitem{HLi2025}
H.H. Li, J.G. Li,  M.R. Xie, W. Zuo, ``Ab initio calculations for shell evolution in proton-dripline nuclei",
\href{https://doi.org/10.1016/j.physletb.2025.139609}
{Phys. Lett. B {\bf 867}, 139609 (2025).}

\bibitem{Gabriel1996}
G. Martínez-Pinedo, A. Poves, E. Caurier, and A. P. Zuker, ``Effective $g_A$ in the $pf$ shell",
\href{https://doi.org/10.1103/PhysRevC.53.R2602}
{Phys. Rev. C {\bf 53}, R2602(R) (1996).}

\bibitem{Yoshida2018}
S. Yoshida, Y. Utsuno, N. Shimizu, and T. Otsuka, ``Systematic shell-model study of $\beta$-decay properties and Gamow-Teller strength distributions in
$A \approx 40$ neutron-rich nuclei",
\href{https://doi.org/10.1103/PhysRevC.97.054321}
{Phys. Rev. C {\bf 97}, 054321 (2018).}

\bibitem{Suhonen2017}
J. Suhonen, ``Value of the Axial-Vector Coupling Strength in $\beta$ and $\beta \beta$ Decays: A Review",
\href{https://doi.org/10.3389/fphy.2017.00055}
{Front. Phys. {\bf 5}, 55 (2017).}

\bibitem{Gysbers2019}
P. Gysbers, G. Hagen, J.D. Holt, G.R. Jansen, T.D. Morris, P. Navrátil, S. Quaglioni, A. Schwenk, S.R. Stroberg and K.A. Wendt, ``Discrepancy between experimental and theoretical $\beta$-decay rates resolved from first principles",
\href{https://doi.org/10.1038/s41567-019-0450-7}
{Nat. Phys. {\bf 15}, 428 (2019).}

\bibitem{Stroberg2021}
S.R. Stroberg, ``Beta Decay in Medium-Mass Nuclei with the In-Medium Similarity Renormalization Group",
\href{https://doi.org/10.3390/particles4040038}
{Particles {\bf 4}, 521 (2021).}

\bibitem{Li2025}
Z. Li, T. Miyagi, A. Schwenk, `` \textit{Ab initio} calculations of beta-decay half-lives for $N=50$ neutron-rich nuclei",
\href{https://doi.org/10.1103/xjv9-t6sn}
{Phys. Rev. Lett. {\bf 136}, 182501 (2026).}


\bibitem{MiyagiMomentsPRL2024}
T. Miyagi, X. Cao, R. Seutin, S. Bacca, R. F. Garcia Ruiz, K. Hebeler, J.D. Holt, and A. Schwenk, ``Impact of Two-Body Currents on Magnetic Dipole Moments of Nuclei",
\href{https://doi.org/10.1103/PhysRevLett.132.232503}
{Phys. Rev. Lett. {\bf 132}, 232503 (2024).}

\bibitem{Brase2026}
C. Brase, T. Miyagi, J. Menéndez, and A. Schwenk, ``Two-body currents at finite momentum transfer and applications to $M1$ transitions",
\href{https://doi.org/10.1103/4tky-t2h1}
{Phys. Rev. C {\bf 113}, 014317 (2026).}



\bibitem{IMSRG_Magnus}
T.D. Morris, N.M. Parzuchowski, and S.K. Bogner, ``Magnus expansion and in-medium similarity renormalization group",
\href{http://dx.doi.org/10.1103/PhysRevC.92.034331}
{Phys. Rev. C {\bf 92}, 034331 (2015).}

\bibitem{EMinteraction}
K. Hebeler, S.K. Bogner, R.J. Furnstahl, A. Nogga, and A. Schwenk, ``Improved nuclear matter calculations from chiral low-momentum interactions",
\href{https://doi.org/10.1103/PhysRevC.83.031301}
{Phys. Rev. C {\bf 83}, 031301(R) (2011).}

\bibitem{NuHamil}
T. Miyagi, ``NuHamil: A numerical code to generate nuclear two- and three-body matrix elements from chiral effective field theory",
\href{https://doi.org/10.1140/epja/s10050-023-01039-y}
{Eur. Phys. J. A {\bf 59}, 150 (2023).}

\bibitem{Miyagi2022PRC}
T. Miyagi, S.R. Stroberg, P. Navrátil, K. Hebeler, and J.D. Holt, ``Converged \textit{ab initio} calculations of heavy nuclei",
\href{https://doi.org/10.1103/PhysRevC.105.014302}
{Phys. Rev. C {\bf 105}, 014302 (2022).}

\bibitem{EM2NPotential}
D.R. Entem and R. Machleidt, ``Accurate charge-dependent nucleon-nucleon potential at fourth order of chiral perturbation theory",
\href{https://doi.org/10.1103/PhysRevC.68.041001}
{Phys. Rev. C {\bf 68}, 041001(R) (2003).}


\bibitem{IMSRGCodeRagnar}
S.R. Stroberg,
\href{https://github.com/ragnarstroberg/imsrg}
{https://github.com/ragnarstroberg/imsrg.}

\bibitem{SuhonenBook}
J. Suhonen, From Nucleons to Nucleus: Concepts of Microscopic
Nuclear Theory, Tensor Operators and the Wigner-Eckart
Theorem (Springer-Verlag, Berlin, 2006), Chap. 2.

\bibitem{Hardy2020}
J.C. Hardy and I.S. Towner, ``Superallowed $0^+ \rightarrow 0^+$ nuclear $\beta$ decays: 2020 critical survey, with implications for $V_{ud}$ and CKM unitarity",
\href{https://doi.org/10.1103/PhysRevC.102.045501}
{Phys. Rev. C {\bf 102}, 045501 (2020).}

\bibitem{Vikas2023}
V. Kumar, P.C. Srivastava, ``Shell-model study of $\beta^+$/EC-decay half-lives for $Z = 21–30$ nuclei", \href{https://doi.org/10.1140/epja/s10050-023-01142-0}
{Eur. Phys. J. A {\bf 59}, 237 (2023).}

\bibitem{Haaranen2017}
M. Haaranen, J. Kotila, and J. Suhonen, ``Spectrum-shape method and the next-to-leading-order terms of the $\beta$-decay shape factor",
\href{https://doi.org/10.1103/PhysRevC.95.024327}
{Phys. Rev. C {\bf 95}, 024327 (2017).}

\bibitem{Haxton2005}
W.C. Haxton, K.M. Nollett, and K.M. Zurek, ``Piecewise moments method: Generalized Lanczos technique for nuclear response surfaces",
\href{https://doi.org/10.1103/PhysRevC.72.065501}
{Phys. Rev. C {\bf 72}, 065501 (2005).}

\bibitem{Caurier2005}
E. Caurier, G. Martínez-Pinedo, F. Nowacki, A. Poves, and A.P. Zuker, ``The shell model as a unified view of nuclear structure",
\href{https://doi.org/10.1103/RevModPhys.77.427}
{Rev. Mod. Phys. {\bf 72}, 427 (2005).}

\bibitem{Utsuno2015}
Y. Utsuno, N. Shimizu, T. Otsuka, S. Ebata, M. Honma, ``Photonuclear reactions of calcium isotopes calculated with the nuclear shell model",
\href{http://dx.doi.org/10.1016/j.pnucene.2014.07.036}
{Prog. Nucl. Energy {\bf 82}, 102 (2015).}


\bibitem{KSHELL} N. Shimizu, T. Mizusaki, Y. Utsuno and Y. Tsunoda, ``Thick-restart block Lanczos method for large-scale shell-model calculations",
\href{https://doi.org/10.1016/j.cpc.2019.06.011}
{Comput. Phys. Comm. {\bf 244}, 372 (2019)}.


\bibitem{DufourZucker1996}
M. Dufour and A.P. Zuker, ``Realistic collective nuclear Hamiltonian",
\href{https://doi.org/10.1103/PhysRevC.54.1641}
{Phys. Rev. C {\bf 54}, 1641 (1996).}


\bibitem{ZuckerPRL2003}
A.P. Zuker, ``Three-Body Monopole Corrections to Realistic Interactions",
\href{https://doi.org/10.1103/PhysRevLett.90.042502}
{Phys. Rev. Lett. {\bf 90}, 042502 (2003).}

\bibitem{NNDC_Q} Q-value calculator (QCalc), National Nuclear Data Center
World Wide Web site,
\href{https://www.nndc.bnl.gov/qcalc/}
{https://www.nndc.bnl.gov/qcalc/}.

\bibitem{kb3g}
A. Poves, J. Sánchez-Solano, E. Caurier, and F. Nowacki, ``Shell model study of the isobaric chains $A=50$, $ A=51$ and $A=52$",
\href{https://doi.org/10.1016/S0375-9474(01)00967-8}
{Nucl. Phys. A {\bf 694}, 157 (2001).}

\bibitem{gxpf1a}
M. Honma, T. Otsuka, B. A. Brown and T. Mizusaki, ``Shell-model description of neutron-rich $pf$-shell nuclei with a new effective interaction GXPF1",
\href{https://doi.org/10.1140/epjad/i2005-06-032-2}
{Eur. Phys. J. A {\bf 25}, 499 (2005).}

\bibitem{IMSRG3f2}
B.C. He and S.R. Stroberg, ``Factorized approximation to the in-medium similarity renormalization group IMSRG(3)",
\href{https://doi.org/10.1103/PhysRevC.110.044317}
{Phys. Rev. C {\bf 110}, 044317 (2024).}

\bibitem{IMSRG3N7_Heinz2021}
M. Heinz, A. Tichai, J. Hoppe, K. Hebeler, and A. Schwenk, ``In-medium similarity renormalization group with three-body operators",
\href{https://doi.org/10.1103/PhysRevC.103.044318}
{Phys. Rev. C {\bf 103}, 044318 (2021).}


\bibitem{IMSRG3N7_Ragnar}
S.R. Stroberg, T.D. Morris, and B.C. He, ``In-medium similarity renormalization group with flowing 3-body operators, and approximations thereof",
\href{https://doi.org/10.1103/PhysRevC.110.044316}
{Phys. Rev. C {\bf 110}, 044316 (2024).}

\bibitem{IMSRG3N7_Heinz2024}
M. Heinz, T. Miyagi, S.R. Stroberg, A. Tichai, K. Hebeler, and A. Schwenk, ``Improved structure of calcium isotopes from \textit{ab initio} calculations",
\href{https://doi.org/10.1103/PhysRevC.111.034311}
{Phys. Rev. C {\bf 111}, 034311 (2025).}

\bibitem{Nichols2014}
A.J. Nichols \textit{et al.}, ``Collectivity in $A \sim 70$ nuclei studied via lifetime measurements in and $^{68,70}$Se'',
\href{http://dx.doi.org/10.1016/j.physletb.2014.04.016}
{Phys. Lett. B {\bf 733}, 52 (2014).}

\bibitem{Kumar2022}
V. Kumar, A. Kumar, P. C. Srivastava, ``Shell-model study for GT-strengths corresponding to $\beta$ decay of $^{60}$Ge and $^{64}$Ge'',
\href{https://doi.org/10.1016/j.nuclphysa.2021.122344}
{Nucl. Phys. A {\bf 1017}, 122344 (2022).}

\bibitem{Stoica2016}
S. Stoica, M. Mirea, O. Niuescu, J.U. Nabi, M. Ishfaq, ``New Phase Space Calculations for $\beta$-Decay Half-Lives",
\href{http://dx.doi.org/10.1155/2016/8729893}
{Adv. High Energy Phys. {\bf 2016}, 8729893 (2016).}








\end{thebibliography}
\end{document}